# Cooperativity and Heterogeneity in Plastic Crystals Studied by Nonlinear Dielectric Spectroscopy

M. Michl, Th. Bauer, P. Lunkenheimer*, and A. Loidl

*Experimental Physics V, Center for Electronic Correlations and Magnetism, University of Augsburg, 86159 Augsburg, Germany*

The glassy dynamics of plastic-crystalline cyclo-octanol and *ortho*-carborane, where only the molecular reorientational degrees of freedom freeze without long-range order, is investigated by nonlinear dielectric spectroscopy. Marked differences to canonical glass formers show up: While molecular cooperativity governs the glassy freezing, it leads to a much weaker slowing down of molecular dynamics than in supercooled liquids. Moreover, the observed nonlinear effects cannot be explained with the same heterogeneity scenario recently applied to canonical glass formers. This supports ideas that molecular relaxation in plastic crystals may be intrinsically non-exponential. Finally, no nonlinear effects were detected for the secondary processes in cyclo-octanol.

As any decent crystalline material, plastic crystals (PCs) exhibit translational symmetry and the molecules as building blocks are fixed on a crystalline lattice. However, this high degree of order does not apply for the orientational degrees of freedom of these molecules, which at high temperatures more or less freely reorient [1,2]. Interestingly, upon cooling these degrees of freedom often freeze into a glasslike, orientationally disordered state and their dynamics exhibits many of the puzzling properties found in canonical (structural) glass formers. Therefore, studying the dynamics of PCs often is considered as an important step on the way to a better understanding of glass-forming liquids and the glass transition in general. Despite a long history of investigation and the general importance of the glassy state of matter, not only for the common silicate glasses but also in such different fields as polymers, metallic glasses, spin glasses, ionic liquids, or biological matter, overall one has to state that this understanding still is poor [3,4,5].

As the molecular motions in PCs are of purely reorientational nature, it seems natural to employ dielectric spectroscopy for their investigation, providing direct access to this type of dynamics. Indeed, investigations of PCs using this method have considerably enhanced our understanding of the glassy state of matter (see, e.g., [2,6,7,8,9,10]). In dielectric spectroscopy, usually the *linear* response of the sample material to an electrical ac field of moderate amplitude is investigated. However, recently the *nonlinear* response of glass-forming matter, detected under high fields of up to several 100 kV/cm, is attracting increasing interest [11,12,13,14,15,16]. Among the pioneering works were dielectric hole-burning experiments, demonstrating that the typical non-exponentiality of glassy dynamics arises from its heterogeneous nature [17]. Further valuable information on this phenomenon was obtained by checking the field-induced variation of the permittivity [11,18]. Moreover, based on a model by Bouchaud, Biroli, and coworkers [19,20], recent measurements of the higher-order susceptibility $\chi_3$ indicate that the non-Arrhenius behavior of glassy dynamics arises from an increase of cooperativity when approaching the glass temperature [13,16].

Non-exponentiality of time-dependence and the non-Arrhenius behavior of the temperature dependence of the molecular dynamics are hallmark features of glass-forming materials and, interestingly, are also found in PCs [2]. However, it is not self-evident that for the latter similar microscopic mechanisms apply as for structural glass formers: For example, it was argued that heterogeneity in PCs may be of less importance than in structural glass formers implying that the dynamics of a single molecule may be intrinsically non-exponential [21,22]. Moreover, the intermolecular interactions leading to cooperativity should be affected by fixing the molecules on lattice positions and indeed the deviations from Arrhenius behavior in PCs are generally weaker than in canonical glass formers [2,23]. Within Angell's strong-fragile classification scheme [24], this implies that PCs are strong glass formers [23], despite exceptions are known [25,26]. Thus, it seems natural to perform nonlinear measurements on PCs, analogous to those on structural glass formers, aiming to clarify the role of heterogeneity and cooperativity in this class of disordered materials. In the present work, we investigate cyclo-octanol, a typical PC [6,27]. Both, the alteration of the dielectric permittivity under high fields and the higher-order susceptibility $\chi_3$ are reported. In addition, results for plastic-crystalline *ortho*-carborane [28,29] are provided.

The experiments were performed using a frequency-response analyzer and the high-voltage boosters "HVB 300" and "HVB 4000" from Novocontrol Technologies. Cyclo-octanol, which is liquid at room temperature, was mixed with 0.1% silica microspheres (2.87 μm average diameter). The sample was put between two highly polished stainless steel plates. The microspheres act as spacing material, leading to a small plate distance that enables the application of high fields of up to 375 kV/cm [30]. For *ortho*-carborane, which is solid at room temperature [29], thin platelets were pressed [31]. To exclude field-induced heating effects, successive high- and low-field measurements were performed, as described in detail in [30]. For cooling, a closed-cycle refrigerator was used.

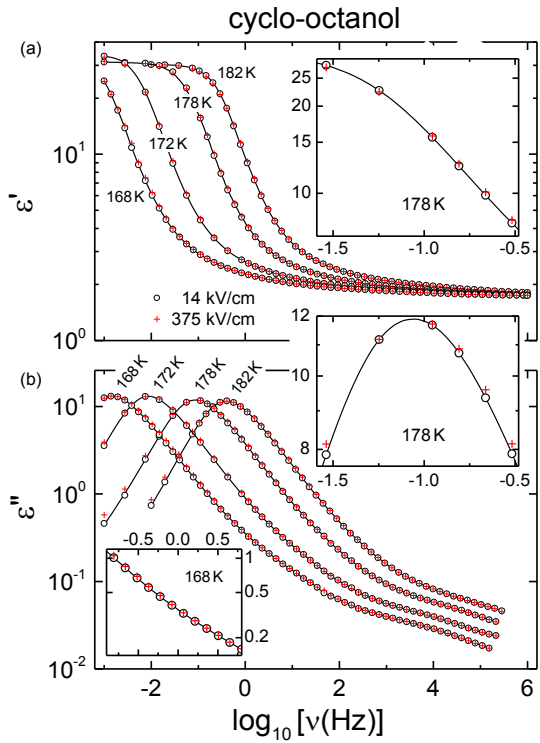

FIG. 1 (color online). Dielectric constant (a) and loss spectra (b) of cyclo-octanol measured at various temperatures and high and low ac fields. The insets provide magnified views illustrating the field-induced variation in the $\alpha$-peak region (right insets) and its absence at higher frequencies (lower left inset). The lines are guides to the eyes.

Figure 1 shows spectra of the real (a) and imaginary part (b) of the dielectric permittivity $\varepsilon^* = \varepsilon' - i\varepsilon''$, obtained at low ($E_l$ = 14.1 kV/cm) and high ac fields ($E_h$ = 375 kV/cm) in the plastic-crystalline phase of cyclo-octanol. To prepare this state, the known transitions to the orientationally ordered state at 265 K and 220 K [33] were supercooled by first cooling the sample to 182 K with a rate of the order of 5 K/min. The subsequent measurements were restricted to temperatures $T_g \approx 168\ \text{K} \leq T \leq 182\ \text{K}$ because at higher temperatures the material tends to order [27]. To collect a sufficient database, the measurements were extended down to relatively low frequencies of $10^{-3}$ Hz.

In Fig. 1 the $\alpha$ relaxation of cyclo-octanol arising from molecular reorientations shows up as step in $\varepsilon'(\nu)$ and peak in $\varepsilon''(\nu)$. They can be fitted [27] by the Cole-Davidson function [34], which well describes the $\alpha$ relaxation of most glass formers [35] and PCs [2]. Moreover, two faster secondary processes, termed $\beta$ and $\gamma$ relaxation [27], are evidenced by shoulders in $\varepsilon''(\nu)$. A closer inspection of Fig. 1 reveals small but significant deviations of the low- and high-field spectra in the region of the $\alpha$ relaxation (see also right insets). Following earlier work [11,15], in Figs. 2(a) and (b) the differences of the high- and low-field spectra are plotted by showing $\Delta \ln \varepsilon = \ln \varepsilon(E_h) - \ln \varepsilon(E_l)$ for $\varepsilon'$ and $\varepsilon''$, respectively. The arrows indicate the $\alpha$-peak positions $\nu_\alpha$ at low fields, related to the $\alpha$-relaxation time via $\tau_\alpha = 1/(2\pi\nu_\alpha)$.

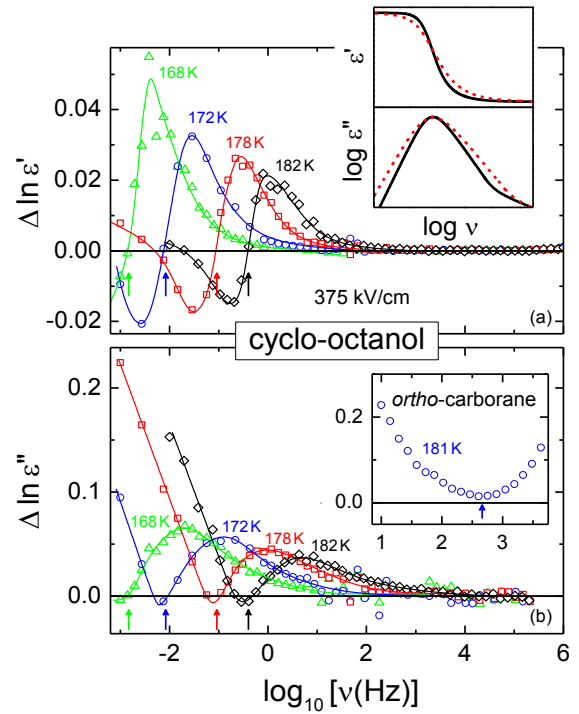

FIG. 2 (color online). Difference of the logarithms of the high- and low-field dielectric spectra of plastic-crystalline cyclo-octanol (cf. Fig. 1), plotted for various temperatures. The arrows indicate the $\alpha$-peak positions [Fig. 1(b)]. The lines are shown to guide the eyes. The upper inset schematically indicates a field-induced broadening (dashed lines: high-field data), which would lead to $\Delta \ln \varepsilon(\nu)$ spectra as observed in the main frames. The lower inset shows $\Delta \ln \varepsilon''$ for *ortho*-carborane at 181 K and 90 kV/cm [31].

Analogous dielectric investigations of canonical glass formers have revealed a lack of nonlinear behavior in the region of the so-called excess wing [15] (but see [36] for results obtained with a modified procedure). This high-frequency spectral feature can be ascribed to a secondary relaxation process [37,38,39]. As seen in the left inset of Fig. 1 and in Fig. 2, in the spectral regions dominated by the secondary relaxations of cyclo-octanol, no field-induced variation of the permittivity is observed and both $\Delta \ln \varepsilon'$ and $\Delta \ln \varepsilon''$ approach zero at high frequencies. This resembles the behavior of the excess-wing relaxation in the canonical glass formers [15]. However, in a recent nonlinear investigation of the $\beta$ relaxation in glass-forming sorbitol [40], nonlinearity indeed was detected. Most likely these results indicate a different nature (intra- *vs.* intermolecular) of the secondary relaxations in both materials. Thus, the use of nonlinear spectroscopy to classify secondary relaxations seems a promising task.

In the $\alpha$-relaxation regime, both $\Delta \ln\varepsilon'$ and $\Delta \ln\varepsilon''$ exhibit peaks at $\nu > \nu_\alpha$. At first glance, this finding resembles the behavior in canonical glass formers [11,15,18], which can be understood within the so-called box model [17,41] assuming a distribution of relaxation times caused by dynamical heterogeneities. Within this scenario, the field-induced increase of $\varepsilon''$ at $\nu > \nu_\alpha$ arises from a selective transfer of field energy into the heterogeneous regions, accelerating their dynamics [11,18]. As the Cole-Davidson function corresponds to a strongly asymmetric relaxation-time distribution [42], only weak absorption should occur for $\nu < \nu_\alpha$ [11] and $\Delta \ln\varepsilon''(\nu)$ should be asymmetric with a minor negative contribution at low frequencies. However, in Fig. 2(b) at $\nu < \nu_\alpha$ we instead find a strong continuous linear increase for decreasing frequency. Moreover, $\Delta \ln\varepsilon'$ at $\nu < \nu_\alpha$ [Fig. 2(a)] assumes negative values, again at variance with the box model as developed for supercooled liquids. Here one should note that, in principle, saturation effects of the polarization can lead to a reduction of the low-frequency dielectric constant, i.e., negative $\Delta \ln\varepsilon'$ [43,44]. However, as found, e.g., in glass-forming 1-propanol [30], in this case a low-frequency plateau in $\Delta \ln\varepsilon'(\nu)$ is expected instead of the minimum revealed by Fig. 2(a).

Remarkably, the results of Fig. 2 are consistent with a simple field-induced broadening of the observed relaxation features as schematically indicated in the upper inset of Fig. 2. While exaggerating the field-induced effects for clarity reasons, the behavior shown in this inset is fully consistent with the experimental data of Fig. 1. Interestingly, the broadening also occurs at the low-frequency flank of the $\alpha$ peak, causing the increase of $\Delta \ln\varepsilon''(\nu)$ at low frequencies [Fig. 2(b)]. Thus, the high-field spectra can no longer be described by the Cole-Davidson function or by the Fourier transform of a stretched-exponential relaxation [45], which predict a linear increase (i.e., a slope one in the log-log plot) at the left flank of the loss peak. The observed peak broadening could indicate stronger disorder and, thus, stronger heterogeneity but it is not clear why a high field should have such an effect. Alternatively, one can assume an intrinsic non-exponentiality, not caused by heterogeneity, in PCs [21,22]. However, it remains to be clarified why this non-exponentiality should increase and deviate from stretched-exponential relaxation under high fields. In any case, it seems clear that this PC exhibits qualitatively different nonlinear behavior than canonical glass formers.

To check if this behavior is typical for PCs, similar measurements in *ortho*-carborane were performed. The lower inset of Fig. 2 reveals qualitatively similar behavior in this PC, namely a field-induced broadening at both flanks of the $\alpha$ peak [31]. Interestingly, for *ortho*-carborane intrinsic non-exponentiality was explicitly reported in [21].

Another important quantity accessible by nonlinear dielectric measurements is the third-order harmonic component of the susceptibility $\chi_3$. In canonical glass formers, $|\chi_3|(\nu)$ was found to show a pronounced hump [13,16]. Within the model by Biroli and coworkers [19,20], it is ascribed to the collective nature of glassy dynamics. (However, it should be noted that a hump in $|\chi_3|(\nu)$ may also arise in the framework of other models [46,47,48].) As revealed by Fig. 3 (showing the dimensionless quantity $|\chi_3|E^2$), $|\chi_3|(\nu)$ of cyclo-octanol also shows a hump. Within the theoretical framework of Refs. [19,20], this finding demonstrates that cooperativity also plays an important role in PCs.

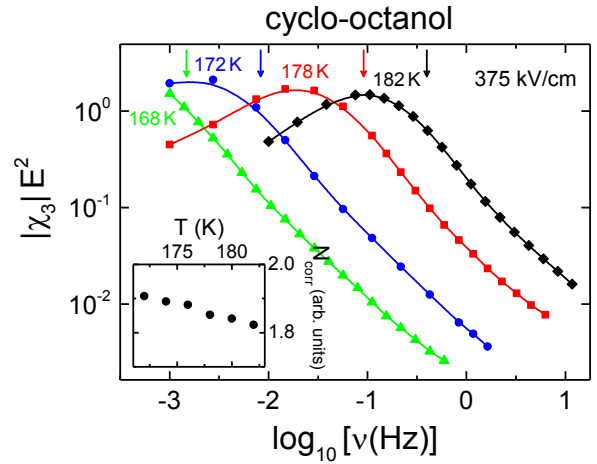

FIG. 3 (color online). Third-order harmonic component of the dielectric susceptibility of plastic-crystalline cyclo-octanol. Spectra of $|\chi_3|E^2$ are shown for various temperatures, measured at a field of 375 kV/cm. The arrows indicate the peak positions in $\varepsilon''(\nu)$ [27]. The lines are guides to the eyes. The inset shows $X_{max} \propto N_{corr}$, deduced from $|\chi_3|$ [13,19,49].

The amplitude of the hump $X_{max}$, showing up in the related quantity $X = |\chi_3| k_B T / [(\Delta\varepsilon)^2 V \varepsilon_0]$, should be proportional to the number of correlated molecules $N_{corr}$ [13,19,49]. Here $\Delta\varepsilon$ is the relaxation strength, $V$ the volume taken up by a single molecule and $\varepsilon_0$ the permittivity of free space. The inset of Fig. 3 shows $N_{corr}$ determined in this way. Obviously, just as for structural glass formers [13,16], the molecular correlations in plastic-crystalline cyclo-octanol increase at low temperatures, even though this increase is rather weak. In [16] a direct proportionality of $N_{corr}$ and the effective energy barrier of the $\alpha$ relaxation was found for various glass-forming liquids. The latter can be estimated from the derivative $H = d(\ln\tau)/d(1/T)$ of the $\alpha$-relaxation time in an Arrhenius plot [4,16]. The lines in Fig. 4(a) show $H(T)$ for cyclo-octanol and for the systems treated in [16] (right scale). Compared to glass-forming glycerol, propylene carbonate

(PCA), and 3-fluoroaniline (FAN), the effective activation energy of cyclo-octanol varies only weakly (the monohydroxy alcohol 2-ethyl-1-hexanol, 2E1H, is a special case, see [16,50]). This implies that the deviations of its $\tau(T)$ from Arrhenius behavior are weaker than in the structural glass formers, in accord with its rather low fragility [6,27], typical for most PCs [2,23]. Within the framework developed in Ref. [16], this finding is in good accord with the detected rather weak variation of $N_{corr}$ (inset of Fig. 3).

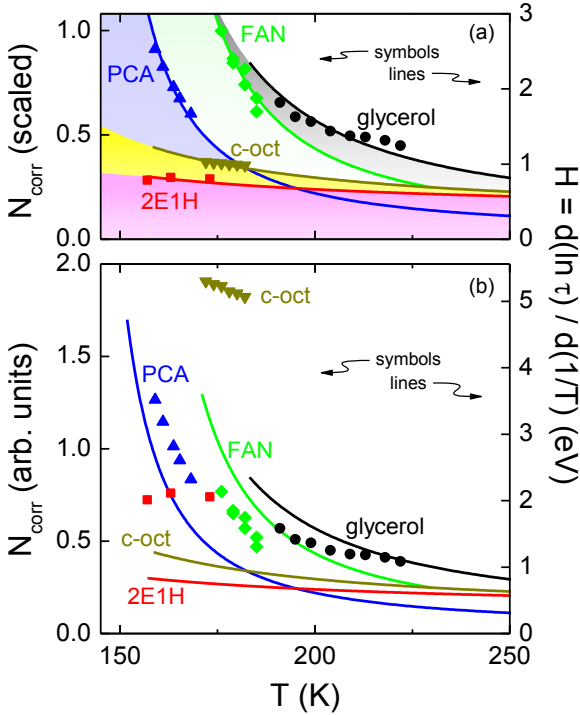

FIG. 4 (color online). The lines show the effective activation energies $H$ of plastic-crystalline cyclo-octanol (present work) and various structural glass formers [16] determined from the derivatives of their temperature-dependent relaxation-times (right scale). For the same materials, the symbols show the number of correlated molecules $N_{corr}$ (left scale). $N_{corr}$ was determined from $\chi_3$ (Fig. 3) and is shown in arbitrary units. In frame (a), the $N_{corr}$ data points were multiplied by separate factors for each material (glycerol: 1.15, PCA: 0.72, FAN: 1.30, 2E1H: 0.39, cyclo-octanol: 0.19) leading to a good match with the derivative curves. In (b), the unscaled data are shown demonstrating the strong deviations for the PC cyclo-octanol.

For a direct comparison, in Fig. 4(a) the results for $N_{corr}$ of cyclo-octanol and the other glass formers [16] are included (left scale). Following Ref. [16], here first the scaling of the $N_{corr}$ ordinate was adjusted to achieve a rough match of $H$ and $N_{corr}$ for glycerol, PCA, and FAN. For each material, an additional scaling factor was applied to $N_{corr}$ to match the effective energy barrier. Indeed, just as for the structural glass formers [16], for cyclo-octanol both quantities can be reasonably scaled onto each other. The much weaker temperature dependence of $N_{corr}$ of the PC, in accord with its weaker variation of $H$, becomes obvious in this figure. One should note that both ordinates in Fig. 4(a) start from zero, implying that the effective activation energy is directly proportional to the number of correlated molecules, i.e. $H = a\, N_{corr}$.

However, there is a significant difference between cyclo-octanol and the structural glass formers, revealed by Fig. 4(b), where the unscaled $N_{corr}(T)$ results are compared to $H(T)$. For glycerol, PCA, and FAN a rough match of $H$ and $N_{corr}$ can be stated even without separate scaling, implying that even the absolute values of $H$ to a large extent are governed by $N_{corr}$ (i.e., the proportionality factors $a$ are similar). This is not the case for 2E1H, which in ref. [16] was ascribed to the formation of string- or ringlike molecular clusters, governing the main relaxation process of many monohydroxy alcohols [50,51]. However, for cyclo-octanol the deviations are even larger, despite the formation of such clusters can be excluded in this PC. Correspondingly, the scaling factor applied in Fig. 4(a) is much smaller than in the other systems (see figure caption). This interesting finding implies that in this PC relatively high values of $N_{corr}$ [52] lead to unexpectedly moderate absolute values of the energy barrier, if compared to the other systems. Obviously, the molecular motion in cyclo-octanol is less impeded by a high $N_{corr}$ than in structural glass formers. In PCs, where the molecules are fixed at a crystalline lattice, molecular correlations are mainly generated by lattice strains, which should reduce the effective energy barrier for the reorientational motion of neighboring molecules. Thus it seems reasonable that in PCs the same $N_{corr}$ leads to lesser impediment of the molecular motions than in supercooled liquids and, thus, smaller effective energy barriers.

In summary, nonlinear dielectric measurements of plastic-crystalline cyclo-octanol at first glance reveal similar behavior as in structural glass formers. In particular, just as for structural glass formers [16], in the plastic-crystalline state it is the temperature variation of molecular cooperativity that causes the non-Arrhenius behavior of glassy dynamics. However, a closer look also reveals several marked differences: For example, in the plastic-crystalline state strong correlations seem to lead to a much weaker slowing down of glassy dynamics than in conventional glass formers. Moreover, an explanation of the field-induced variation of the permittivity of cyclo-octanol and *ortho*-carborane within the heterogeneity scenario as developed for supercooled liquids is not straightforward, which may indicate that molecular motions in PCs are intrinsically non-exponential [21,22,53]. Finally, we find no nonlinear behavior of the secondary relaxations of cyclo-octanol. Thus, while sharing many characteristics of glassy dynamics with canonical glass formers, on a microscopic level PCs may behave quite different, shedding new light on cooperativity and heterogeneity in glassy systems in general.

We thank R. Richert for helpful discussions. This work was supported by the Deutsche Forschungsgemeinschaft via Research Unit FOR1394.

\* Corresponding author.
Peter.Lunkenheimer@Physik.Uni-Augsburg.de

# Cooperativity and Heterogeneity in Plastic Crystals Studied by Nonlinear Dielectric Spectroscopy

## Supplemental Material

M. Michl, Th. Bauer, P. Lunkenheimer*, and A. Loidl

*Experimental Physics V, Center for Electronic Correlations and Magnetism, University of Augsburg, 86159 Augsburg, Germany*

Here we provide more information on the nonlinear measurements of *ortho*-carborane. As this plastic crystal (PC) is solid at room temperature, for the measurements thin platelets were pressed between polished stainless-steel plates serving as electrodes. Due to the limited stability of the pressed carborane plates, their minimum thickness was of the order of 100 μm, much larger than the plate distance for cyclo-octanol (about 3 μm), which is liquid at room temperature enabling the use of microspheres for plate separation. To achieve sufficiently high electrical fields despite the higher thickness, a "HVB 4000" high-voltage booster (Novocontrol Technologies) was used. Compared to the "HVB 300" employed for the measurements of cyclo-octanol, it enables applying higher voltages, however, within a smaller frequency range only. In this way, fields of about 90 kV/cm could be reached before electrical breakdown occurred. Due to uncertainties of the thickness determination, the this field has an ucertainty of about ± 20 kV/cm. For temperatures, where the relaxation process of *ortho*-carborane is within the available frequency window (about 10 Hz - 10 kHz), we found this material to transfer to the completely (translationally *and* orientationally) ordered crystalline state if cooled slowly [1]. Thus, the frequency sweeps were performed directly after quickly cooling the sample to the desired measurement temperature.

Figure S1(a) shows the loss spectra for low and high fields measured at two temperatures. Just as for cyclo-octanol, a field-induced enhancement of $\varepsilon''$ is observed at both, the left and right flank of the relaxation peak. In Fig. S1(b) the logarithmic difference $\Delta \ln \varepsilon''$ is plotted. Similar to the findings in cyclo-octanol (Fig. 2 of main article), this quantity increases at frequencies below and above the loss-peak frequency, which is indicated by the arrows. Notably, the field-induced variation of the loss in ortho-carborane seems to be even stronger than in cyclo-octanol.

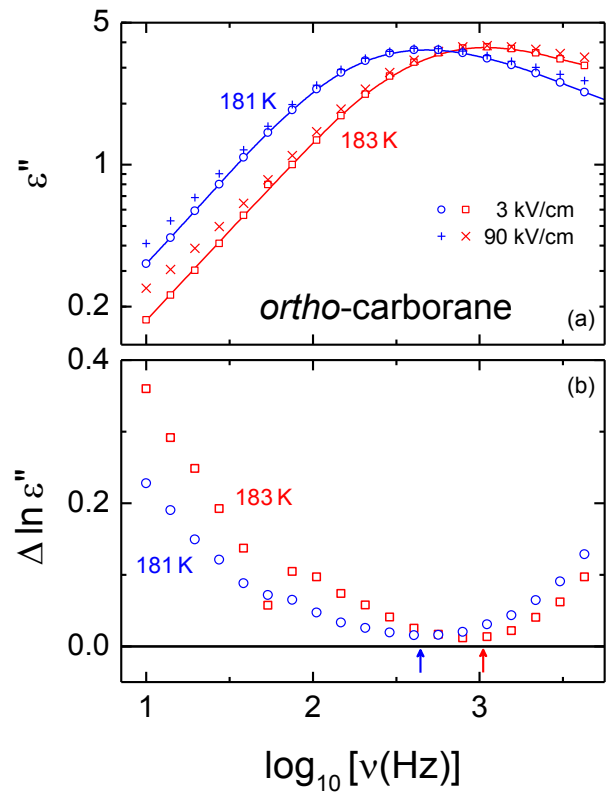

FIG. S1. (a) Dielectric loss spectra of *ortho*-carborane in the plastic-crystalline state for two temperatures, measured at low and high field. The lines are shown to guide the eyes. (b) Difference of the logarithms of the high- and low-field spectra. The arrows indicate the loss-peak positions.